\documentclass[12pt]{article}

\usepackage{amsmath,amssymb,amsfonts,xspace}
\usepackage[latin1]{inputenc}
\usepackage{fullpage}

\numberwithin{equation}{section}

\newcommand{\N}{{\mathcal N}}
\newcommand{\C}{{\mathbb C}}
\newcommand{\cC}{{\mathcal C}}
\newcommand{\cM}{{\mathcal M}}
\newcommand{\PP}{{\mathbb P}}
\newcommand{\R}{{\mathbb R}}
\newcommand{\HH}{{\mathbb H}}

\newcommand{\cO}{\mathcal{O}}

\newcommand{\A}{{\mathbb A}}
\newcommand{\p}{{\partial}}
\newcommand{\eps}{{\epsilon}}

\newcommand{\kahler}{{K\"ahler}\xspace}
\newcommand{\hk}{{hyperk\"ahler}\xspace}

\begin{document}

\thispagestyle{empty}
\addtocounter{page}{-1}


\vskip-0.35cm
\begin{flushright}
YITP-SB-07-42
\end{flushright}
\vspace*{0.2cm}
\centerline{\LARGE \bf A note on conformal symmetry}
\centerline{\LARGE \bf in projective superspace}
\vspace*{1.0cm}
\centerline{\bf Radu A. Iona\c{s}$\,^a$ and Andrew Neitzke$\,^b$}
\vspace*{0.7cm}
\centerline{\it $^a$ C.N.Yang Institute for Theoretical Physics, Stony Brook University}
\centerline{\it Stony Brook, NY 11794-3840, USA }
\centerline{\tt ionas@max2.physics.sunysb.edu}
\vspace*{0.2cm}
\centerline{\it $^b$ Jefferson Physical Laboratory, Harvard University}
\centerline{\it Cambridge, MA 02138, USA}
\centerline{\tt neitzke@physics.harvard.edu}

\vspace*{0.8cm}
\begin{abstract}
\noindent We describe a sufficient condition for actions constructed in projective superspace to possess an
$SU(2)$ R-symmetry.   We check directly that this condition implies that the corresponding
hyperk\"ahler varieties, constructed by means of the generalized Legendre transform, have a Swann bundle structure.
\end{abstract}
\newpage

\section{Introduction}

\subsection*{Motivation}

The projective superspace formalism \cite{Lindstrom:1987ks} has proven useful in
constructing field theories which realize $\N=2$, $d=4$ supersymmetry off-shell.
Indeed, for theories admitting such a construction, projective superspace
generally gives the most economical description of the action.
It is therefore of interest to know how this formalism can
be coupled to supergravity.

A standard approach to constructing such a coupling would be to couple a rigid theory to superconformal gravity
and then break the superconformal symmetry by gauge-fixing.  In order to couple to superconformal gravity,
though, the rigid action $S$ must possess a superconformal symmetry.  In particular, there must be an action of
$SU(2) \times \R^\times$ on the field space, where $SU(2)$ is an R-symmetry preserving $S$, and $\lambda \in
\R^\times$ rescales the action, $S \mapsto \lambda^2 S$.

In this note we discuss some conditions under which an action constructed in projective superspace has this
$SU(2) \times \R^\times$ symmetry.  The condition we propose is simple and presumably known to experts
(see in particular \cite{Kuzenko:2007qy,Alexandrov:2008nk}.)
In the first part of the note we describe it in a coordinate-free way which makes its origin particularly
transparent (at least to us).  In the second part we discuss the consequences of this
$SU(2) \times \R^\times$ symmetry for theories which are realized as nonlinear sigma models.

\subsection*{The condition}

In the projective superspace formalism, the action is obtained by integrating a superfield $G$ defined
on ``analytic superspace'' $\A$, with bosonic part
$\R^4 \times \C\PP^1$.  The integral runs over a submanifold $\cC \subset \A$ which projects to a closed contour
in $\C\PP^1$ and is extended along all of the fermionic directions.

We are interested in constructing an $SU(2)_R$-invariant action.  The fermionic directions in $\A$ admit
an $SU(2)_R$-invariant measure $\mu_f$ valued in the line bundle $\cO(-4)$ over $\C\PP^1$,
while $\C\PP^1$ has an $SU(2)_R$-invariant holomorphic 1-form $\mu_b$ valued
in $\cO(2)$.  Hence if we write
\begin{equation}
\mu = \mu_b \times \mu_f
\end{equation}
then $\mu$ is a $SU(2)_R$-invariant measure valued in $\cO(-2)$.

If $G$ is a section of $\cO(2)$, then,
the action
\begin{equation}
S = \int d^4x \oint_\cC \mu\,G
\end{equation}
has the desired $SU(2)_R$ invariance.  This is our criterion.

\subsection*{Geometric consequences}

One may further assume that the action density superfield $G$ is constructed from one or several ``${\cal O}(2j)$ multiplets.''
These multiplets are represented by superfields on $\A$ which are sections of the bundles ${\cal O}(2j)$ over $\mathbb{CP}^1$, satisfying a reality condition induced by the antipodal conjugation on the Riemann sphere.
From an ${\cal N}=1$ point of view, these multiplets consist of a chiral superfield, a complex linear superfield and a number of auxiliary fields.

Relaxing the constraints on the linear superfields by means of a chiral Lagrange multiplier and then integrating out the unconstrained superfields yields a dual action involving only chiral superfields.  Such an action generally defines a nonlinear
sigma model on a \kahler manifold $\cM$.
In fact, in this case $\cM$ must be \hk, because of the $\N=2$ supersymmetry.  This construction is the superspace equivalent of Hodge duality between $0$-form and $2$-form gauge fields in four dimensions.  It yields a K\"ahler potential for $\cM$ as a generalized Legendre transform of a contour integral of $G$, and a set of holomorphic coordinates on $\cM$.  Discovered originally in a supersymmetric field-theoretic context \cite{Lindstrom:1983rt}, this construction was later described geometrically in terms of the twistor space of $\cM$ \cite{Hitchin:1986ea,Ivanov:1995cy} and has been used
extensively to construct \hk manifolds, see e.g. \cite{Hitchin:1986ea,Ivanov:1995cy,deWit:2001dj,Arai:2006gg}.

As we discussed above, if $G$ is a section of $\cO(2)$, then our action $S$ has an $SU(2) \times \R^\times$ symmetry. One may then ask:  what structure does this symmetry imply for the
\hk space $\cM$?  We show that $\cM$ is a Swann bundle, i.e. it has an action of
the nonzero quaternions $\HH^\times$, whose real component acts homothetically while the three purely imaginary components act isometrically.  Swann bundles have the remarkable feature that they admit a function $\chi$ which
is simultaneously a \kahler potential for all of the standard 2-sphere's worth of complex structures.
The generalized Legendre transform construction, which generally yields a \kahler potential only for
one complex structure, gives in the case of Swann bundles precisely this $\chi$.

\subsection*{Prior work}

Our discussion generalizes several results of \cite{deWit:2001dj}:  there it was shown that
if $G$ is a function of superfields $\eta_I$, all of which are sections of $\cO(2)$, and moreover if
$G(\{\eta_I\})$ is homogeneous of degree $1$,
then $S[G]$ has an $SU(2)_R$ symmetry and the hyperk\"ahler variety constructed by means of the generalized Legendre transform from $G$ is a Swann bundle.

While this paper was in preparation, the papers \cite{Kuzenko:2007qy,Alexandrov:2008nk}
appeared discussing essentially the same homogeneity condition which we
consider, and showing that it gives superconformal actions.
In addition, the important problem of coupling such superconformal actions to conformal supergravity
has recently been treated in \cite{Kuzenko:2008ep}.

\subsection*{Organization}

In Section \ref{SEC:analytic-supersp} of this note we review the concept of analytic superspace.
In Section \ref{SEC:global-arg} we explain in more detail the ${\cal O}(2)$ condition sketched above and then demonstrate how the same criterion arises if one works in fixed local coordinates on $\C \subset \C\PP^1$ and a fixed trivialization of the line bundles, as is usually done in the literature on the projective superspace formalism.
In Section \ref{SEC:O(2j)} we outline the properties of the particular class of ${\cal O}(2j)$ analytic superspace multiplets.
In Section \ref{SEC:scalar-tensor} we recall the details of scalar-tensor duality and of the generalized Legendre transform approach to constructing hyperk\"ahler metrics.
In Section \ref{SEC:HKC-from-GLT} we investigate the local properties of the metrics derived from meromorphic potentials $G$ which are functions of ${\cal O}(2j)$ multiplets and satisfy the ${\cal O}(2)$ criterion.
In section \ref{SEC:HKC} we review generic properties of Swann bundles and in section \ref{SEC:App} we discuss briefly several applications.

\section{Analytic superspace} \label{SEC:analytic-supersp}

We begin by fixing some notation for $\R^{3,1|8}$.
The Grassman-odd derivative operators along the eight fermionic directions are $D_{A\alpha}$ and their conjugates $\bar{D}^A_{\dot\alpha}$; here $A \in \{1,2\}$ is the $R$-symmetry index and $\alpha$, $\dot\alpha$ are complex
2-component spinor indices.  These operators obey
\begin{align} \label{N=2_Ds}
\{D_{A\alpha}, D_{B\beta}\} = 0, && \{D_{A\alpha}, \bar{D}^{B}_{\dot\alpha}\} = i \delta_{A}{}^{B} \p_{\alpha \dot\alpha}.
\end{align}

Let $\pi^A$ be complex coordinates on an auxiliary $\C^2$ transforming in the doublet of $SU(2)_R$.
We consider them as homogeneous coordinates for $\C\PP^1$.
Then a meromorphic function $f(\pi^A)$ which is homogeneous
of degree zero is a meromorphic function on $\C\PP^1$.  More generally,
meromorphic $f(\pi^A)$ which are homogeneous of nonzero degree $k$ are
meromorphic sections of the line bundle $\cO(k)$ over $\C\PP^1$.  We will use this
terminology repeatedly in what follows.

The analytic superspace $\A$ is an extension of Minkowski space $\R^{3,1}$ by an auxiliary $\C\PP^1$,
with four fermionic directions fibered over it.  It is most naturally defined as a quotient of
$\R^{3,1|8} \times \C\PP^1$ by the odd $\cO(1)$-valued vector fields
\begin{align} \label{def-nabla}
\nabla = \pi^1 D_1 + \pi^2 D_2 = \pi^A D_A, && \bar\nabla = -\pi^2 \bar{D}^1 + \pi^1 \bar{D}^2 = \pi_A \bar{D}^A.
\end{align}
In \eqref{def-nabla} the $R$-symmetry index is lowered by means of the two-dimensional $\epsilon$-tensor,
$\pi_A = \pi^B\epsilon_{BA}$.  Here and henceforward we suppress the spectator spinor indices $\alpha$ and $\dot\alpha$
wherever possible.

From (\ref{N=2_Ds}) it follows that
\begin{equation} \label{nabla-alg}
\{\nabla,\nabla\} = \{\nabla,\bar{\nabla}\} = \{\bar{\nabla},\bar{\nabla}\} = 0.
\end{equation}
Hence the vector fields $\nabla_\alpha$ and $\bar\nabla_{\dot\alpha}$ generate a
$(0|4)$-dimensional abelian supergroup.  It follows that the quotient $\A$ has dimension
$(5|8)-(0|4)=(5|4)$.
Concretely, superfields on $\A$ are functions on $\R^{3,1|8} \times \C\PP^1$ which are
simultaneously annihilated by all $\nabla_\alpha$ and $\bar\nabla_{\dot\alpha}$.

\section{Building $SU(2)_R$ and $\N=2$ invariant actions} \label{SEC:global-arg}

Now we want to construct an action invariant under $\N=2$ supersymmetry and $SU(2)_R$.
Following the standard recipe, we will define
the action by integrating superfields over $\A$, and
realize the supersymmetry variations as shifts along the fermionic directions.

More specifically, we are interested in obtaining an $SU(2)_R$-invariant
action, so we should realize the $SU(2)_R$ symmetry directly on $\A$.
Indeed, there is a natural action of $SU(2)$ on $\R^{3,1|8} \times \C^2$, obtained by combining the R-symmetry
action on the fermionic derivatives $D_A$, $\bar{D}^A$ with the standard linear action on the $\pi^A$.
Both $\nabla$ and $\bar\nabla$ are invariant under this action, and it also commutes with the $\C^\times$
action by overall rescaling on $\C^2$; therefore it descends to give an $SU(2)$ action on $\A$, and also
on the line bundles $\cO(k)$ over $\A$.

We begin by discussing the appropriate $SU(2)$-invariant measure for integration on $\A$.

\subsection*{The bosonic measure}

First we consider the bosonic directions
of $\C\PP^1$.  We want to construct our action as a contour integral, so specifying a measure here just means
specifying a holomorphic 1-form.
However, it is well known that $\C\PP^1$ does not possess any $SU(2)$-invariant holomorphic 1-form
(nor indeed any holomorphic 1-form at all) in the usual sense.  The closest we can get is to write an
$SU(2)$-invariant holomorphic 1-form valued in the line bundle $\cO(2)$, namely
\begin{equation} \label{mu_b}
\mu_b = \pi_A d\pi^A.
\end{equation}

\subsection*{The fermionic measure}

Next we consider the four fermionic directions.  To construct an $SU(2)$-invariant measure here
we introduce operators $\Delta_\alpha$, $\bar\Delta_{\dot\alpha}$, defined by
\begin{align}
\Delta = \lambda^A(\pi) D_A, && \bar{\Delta} = \lambda_A(\pi) \bar{D}^A, \label{delta-ops}
\end{align}
where we choose the functions $\lambda^A(\pi)$ subject to the constraint
\begin{equation} \label{lambda-constraint}
\lambda_A(\pi) \pi^A = 2.
\end{equation}
This constraint does not determine $\lambda^A(\pi)$ uniquely.  However, the freedom in choosing
$\lambda^A(\pi)$ does not lead to an ambiguity in the operators $\Delta$
and $\bar\Delta$:  the reason is that changing $\lambda^A$ is equivalent to redefining $\Delta$ by a multiple
of $\nabla$, but $\nabla = 0$ when acting on superfields on $\A$.  So \eqref{delta-ops} and
\eqref{lambda-constraint} define $\Delta$, $\bar\Delta$ uniquely.  In particular, since \eqref{delta-ops} and \eqref{lambda-constraint} are $SU(2)$ invariant it follows that $\Delta$, $\bar\Delta$
are $SU(2)$ invariant as well.  From \eqref{lambda-constraint} we
see that they have weight $-1$ under rescaling of the $\pi^A$, or in other words they are $\cO(-1)$-valued.

Writing $\Delta^2 = \eps^{\alpha\beta}\Delta_\alpha\Delta_\beta$, and likewise $\bar\Delta^2$,
we get the $SU(2)$-invariant and $\cO(-4)$-valued measure for integration over the fermions,
\begin{equation} \label{mu_f}
\mu_f = \Delta^2 \bar\Delta^2.
\end{equation}
This measure can be rewritten in various equivalent ways.  For example, denoting $D_1 = {\cal D}$, one can write
\begin{align}
\Delta = \frac{1}{\pi^2} (2{\cal D} + \lambda^2 \nabla), && \bar{\Delta} = \frac{1}{\pi^1} (2\bar{\cal D} + \lambda^1 \bar{\nabla})
\end{align}
so
\begin{equation} \label{mu_f_specialized}
\mu_f = \Delta^2\bar{\Delta}^2  = \left(\!\frac{4}{\pi^1\pi^2}\!\right)^{\!2} \!{\cal D}^2\bar{\cal D}^2 + \mbox{trivial terms},
\end{equation}
where by ``trivial terms'' we mean either terms proportional to $\nabla$ or $\bar{\nabla}$ (which vanish when acting on superfields defined on $\A$) or terms proportional to space-time derivatives (which vanish upon integration over space-time with appropriate boundary conditions.)

\subsection*{Constructing the action}


Combining the measures for integration over bosons and fermions,
\begin{equation}
\mu = \mu_b \times \mu_f
\end{equation}
is an $SU(2)$-invariant measure, valued in the line bundle $\cO(-2)$.

Now we can explain the construction of invariant actions.  Suppose we are given some collection
of superfields $\eta_I$ over $\A$
and a composite ``action density'' superfield $G(\eta_I)$.  Suppose moreover that $G(\eta_I)$
is a section of $\cO(2)$.  Then multiplying
$\mu$ by $G(\eta_I)$ gives a measure valued in $\cO(0)$, i.e. a measure in the
standard sense.
So choose a contour $\gamma \subset \C\PP^1$
and define $\cC \subset \A$ to be the inverse image
of $\gamma$ under the projection $\A \to \C\PP^1$.
Integrating over $\cC$ means integrating over $\gamma$ and the four fermionic directions
fibered over it.  We define the action as
\begin{equation} \label{tensor-action}
S[\{\eta_I\}] = \int d^4x \oint_\cC \mu\,G(\eta_I).
\end{equation}

\subsection*{$\N=2$ supersymmetry invariance}

The action \eqref{tensor-action} is invariant under $\N = 2$ supersymmetry, because the supersymmetry
variations acting on $G$ just produce total derivatives in superspace.
More explicitly, this can be seen as follows: the infinitesimal variation of $G$ is
\begin{equation}
\delta G = (\epsilon^{\alpha A} Q_{\alpha A} + \bar{\epsilon}^{\dot{\alpha}}_A \bar{Q}_{\dot{\alpha}}^A)G.
\end{equation}
The supersymmetry generators $Q_{\alpha A}$ and $\bar{Q}^A_{\dot{\alpha}}$ differ from the corresponding superderivatives $D_{\alpha A}$ and $\bar{D}^A_{\dot{\alpha}}$ only by total space-time derivatives; consequently, under the action integral we may replace $Q \to D$, $\bar{Q} \to \bar{D}$. The superderivatives, in
turn, can be replaced by linear combinations of $\nabla$ and $\Delta$ (respectively $\bar{\nabla}$
and $\bar{\Delta}$).  The terms containing $\nabla$ and $\bar{\nabla}$
vanish when acting on $G$, by the definition of $\A$. The terms containing $\Delta$ and
$\bar{\Delta}$ are annihilated by the fermionic measure using the identities $\Delta^3 = \bar{\Delta}^3 = 0$.

\subsection*{$SU(2)$ invariance}

As we explained above, if $G$ is a section of $\cO(2)$, then the action $\mu G$ is
$SU(2)$ invariant.
We would like to conclude from this that $S$
is invariant under the simultaneous action of $SU(2)$ on all $\eta_I$.
This will be so if $\gamma$ is chosen in an $SU(2)$ covariant way.

For example, at least locally in field space, we can choose $\gamma$ to be a
closed contour encircling some fixed set of singularities
of $G(\eta_I)$.  Then $S$ is invariant under \textit{any} infinitesimal deformations of $\gamma$, and in
particular it is invariant under the deformation of $\gamma$ arising from an infinitesimal $SU(2)$
rotation.   Combined with the $SU(2)$ invariance of $\mu$ this allows us
to conclude that $S$ is invariant at least under the Lie algebra ${\mathfrak{su}}(2)$.

Another possibility is to choose $\gamma$ to be an open contour with endpoints determined by
some algebraic equation in the $\eta_I$ (e.g. $\eta_I = 0$); then the $SU(2)$ action on the $\eta_I$ will
transform the endpoints of $\gamma$ according to the standard action on $\C\PP^1$ (see also
\cite{ell-hk} for a more detailed discussion.)  Combining this with the
$SU(2)$ invariance of $\mu$ we again conclude that $S$ is $SU(2)$ invariant.

\subsection*{${\mathbb R}^\times$ symmetry}

In addition to the $SU(2)$ preserving $S$, we would like to have an $\R^\times$ action under
which $S$ is rescaled, $S \mapsto \lambda^2 S$.  If $G(\eta_I)$ is a section of $\cO(2)$ as considered above, then
this action is easy to obtain:  we just consider the overall rescaling $\pi^A \mapsto \lambda \pi^A$ for
$\lambda \in \R$.  This action transforms the superfields $\eta_I$, in such a way that $G(\eta_I)$
transforms as desired.

\subsection*{Local coordinates}

For practical computations, as well as to recover the standard formulas in the literature, we should translate
the prescriptions given above into local coordinates on $\C\PP^1$.  For this purpose we choose the inhomogeneous coordinate
\begin{equation}
\zeta = \frac{\pi^2}{\pi^1}.
\end{equation}
We also fix the $\C^\times$ rescaling freedom by evaluating all superfields on the locus
\begin{equation}
\pi^1 \pi^2 = 1 \label{locus}
\end{equation}
or in other words, setting
\begin{equation}
\pi^1 = \frac{1}{\sqrt{\zeta}}, \qquad \pi^2 = \sqrt{\zeta}
\end{equation}
(The sign ambiguity in the square roots is of no consequence as long as we consider $\cO(k)$ only for $k$ even.)

In this trivialization the bosonic measure \eqref{mu_b} becomes
\begin{equation}
\mu_b = \frac{d \zeta}{\zeta}
\end{equation}
and using \eqref{mu_f_specialized} we can write the fermionic measure (modulo trivial terms) as
\begin{equation}
\mu_f = 16\, {\cal D}^2\bar{\cal D}^2.
\end{equation}
The action \eqref{tensor-action} then becomes
\begin{equation} \label{N=1-action}
S = \int d^4x\, {\cal D}^2\bar{\cal D}^2\! \oint_\gamma \frac{d\zeta}{\zeta} G(\eta).
\end{equation}
In this presentation only $\N=1$ supersymmetry is manifest.

In \eqref{N=1-action} there appears to be a pole at $\zeta = 0$.
We emphasize that all our constructions are $SU(2)$ invariant and hence the point $\zeta = 0$ is not
special in any way; indeed, the pole could have been moved to any other $\zeta$ by a different
choice of local coordinates and trivializations.

To make \eqref{N=1-action} more explicit we should describe how the superfields
decompose under the $\N=1$ subalgebra generated by ${\cal D}$, ${\cal \bar D}$.  We discuss this in the next section.

\section{${\cal O}(2j)$ multiplets} \label{SEC:O(2j)}

As mentioned in the introduction, we will pay particular attention to a class of analytic supermultiplets $\eta_I$, the  $\cO(2j)$ multiplets (with $j$ a positive integer) \cite{Lindstrom:1987ks}. These are sections
\begin{equation}
\eta^{(2j)} = \eta_{A_1\cdots A_{2j}}\pi^{A_1} \cdots \pi^{A_{2j}}
\end{equation}
of $\cO(2j)$ over $\A$,
satisfying an additional reality condition defined using the $SU(2)$-invariant real structure $\eps_{AB}$ on $\C^2$,
\begin{equation} \label{reality-constr}
\eta^{(2j)}(\pi^1, \pi^2) = \overline{\eta^{(2j)}(-\bar\pi^2, \bar\pi^1)}.
\end{equation}
Each of the $2j+1$ components $\eta_{A_1 \cdots A_{2j}}$ is a superfield on $\R^{3,1 \vert 8}$; they are related
by the analytic superspace constraints
\begin{equation}
\nabla \eta^{(2j)} = \bar{\nabla} \eta^{(2j)} = 0
\end{equation}
which require
\begin{equation}
D_{(A} \eta_{A_1 \cdots A_{2j})} = 0. \label{eta-N=2-constr}
\end{equation}
In our local coordinates and trivialization such an $\eta$ is written
\begin{equation}
\eta^{(2j)}(\zeta)=\frac{\bar{z}}{\zeta^j} + \frac{\bar{v}}{\zeta^{j-1}} + \frac{\bar{t}}{\zeta^{j-2}} + \cdots + x + (-)^j(\cdots + t \zeta^{j-2} - v\zeta^{j-1} + z \zeta^j),
\end{equation}
where the coefficients come in complex conjugate pairs, except for the middle coefficient, $x$, which is real.

From an ${\cal N}=1$ superspace point of view only the first two components of the multiplet, $z$ and $v$, are constrained
(along with their conjugates $\bar{z}$ and $\bar{v}$): from \eqref{eta-N=2-constr}, we get
\begin{align}
{\cal D} \bar{z} & =  0 \label{z-chiral} \\[5pt]
{\cal D}^2 \bar{v} & = 0 \label{v-linear}.
\end{align}
Thus $\bar{z}$ is an anti-chiral superfield and $\bar{v}$ is a complex linear superfield. The rest of the components of the multiplet are unconstrained in $\N = 1$ superspace; in other words, they are auxiliary superfields.

\subsection*{An extension}

So far we have argued that if $G$ is $\cO(2)$-valued then $S$ constructed in \eqref{tensor-action} is $SU(2)$ invariant.
In fact, we can allow something slightly more general, as we now describe.

Observe first that $G = \eta^{(2)}$ is $\cO(2)$-valued, but
the corresponding action \eqref{tensor-action} vanishes.  That is because $\eta^{(2)}$ has
only three components --- a chiral superfield, its complex
conjugate and a real linear superfield --- all of which are annihilated by the $\N = 1$ superspace measure in \eqref{N=1-action}, as follows from \eqref{z-chiral} and \eqref{v-linear}.  One can also understand this statement
geometrically:  since both $\eta^{(2)}$ and $\mu$ are regular (albeit twisted by $\cO(2)$ and $\cO(-2)$ respectively),
the product $\mu \eta^{(2)}$ must also be regular, so integrating it over the fermionic directions must give a
holomorphic 1-form on $\C\PP^1$; but the only such 1-form is zero.

Next we consider
\begin{equation}
G = \eta^{(2)} \log \eta^{(l)}
\end{equation}
where $l$ is any even integer.
Because of the logarithm $G$ is strictly speaking not well defined:  it suffers from an ambiguous choice of branch.
Nevertheless, its integral over the fermionic directions is well defined, because the ambiguity is just a multiple
of $\eta^{(2)}$, which vanishes after integration.  Hence the integral of $\mu G$ over the fermions is a well defined
holomorphic 1-form on $\C\PP^1$.  Moreover, the presence of the extra logarithm does not
destroy the $SU(2) \times \R^\times$ symmetry we have discussed.  Indeed, upon choosing any
fixed gauge for the $\C^\times$ symmetry, one sees that under an $SU(2)$
transformation, $\eta^{(l)}$ transforms like a scalar except for a rescaling by some function
$f(\zeta)^l$.  As a consequence $\log \eta^{(l)}$ transforms like a scalar except for a constant shift by
$l \log f(\zeta)$; this extra shift corresponds to shifting $G$ by $\eta^{(2)} (l \log f(\zeta))$, which vanishes
after integration over the fermions, as discussed above.

\section{Scalar-tensor duality and the generalized Legendre transform construction} \label{SEC:scalar-tensor}

Let us now consider the class of action densities $G$ that depend exclusively on multiplets $\eta_I$ of ${\cal O}(2j_I)$ type. At this point we do not assume that these action densities satisfy the ${\cal O}(2)$ condition, they are rather generic. To avoid notational clutter, we shall work in effect with functions $G$ depending on only one ${\cal O}(2j)$ multiplet, which we denote simply by $\eta$. We stress that this limitation is dictated only  by practical reasons; the arguments and results to follow can be straightforwardly generalized to include
several such multiplets or combinations of multiplets with different values of $j$, including $j=1$, just by introducing additional indices distinguishing the various multiplets and summing over.

The theories governed by this type of actions turn out to have dual formulations in terms of ${\cal N}=2$ supersymmetric sigma-models whose hyperk\"ahler target spaces admit certain Killing spinors \cite{Lindstrom:2008hx}. This duality relation, termed ``scalar-tensor duality" for reasons that will become clear shortly,  was discovered originally in \cite{Lindstrom:1983rt} and extended to the present context in \cite{Lindstrom:1987ks}.

To describe it, we start with the reduced ${\cal N}=1$ form \eqref{N=1-action} of the otherwise ${\cal N}=2$ action for $\eta$,\footnote{Note that since we do not assume here that $G$ is an ${\cal O}(2)$ section over $\A$, this action need not be $SU(2)$-invariant.}
\begin{equation}
S = \int d^4 x {\cal D}^2\bar{\cal D}^2 F(z,\bar{z},v,\bar{v},t,\bar{t}, \cdots \!, x),
\end{equation}
with the function in the integrand given by
\begin{equation}
F(z,\bar{z},v,\bar{v},t,\bar{t}, \cdots \!, x) = \oint_{\gamma} \frac{d\zeta}{\zeta} G(\eta(\zeta)). \label{F-pot}
\end{equation}
One can relax the constraint on the linear superfield $v$ by means of a chiral Lagrange multiplier $u$. This yields the first-order action
\begin{equation}
S_1 = \int d^4 x {\cal D}^2\bar{\cal D}^2 [F(z,\bar{z},v,\bar{v},t,\bar{t}, \cdots \!, x) - uv - \bar{u}\bar{v}].
\end{equation}
Indeed, by varying\footnote{The variation of the action with respect to a chiral superfield should respect the chiral constraint. $\bar{\cal D} u = 0$ is solved by $u = \bar{\cal D}^2 \psi$, with $\psi$ an unconstrained superfield. The chiral constraint is automatically preserved if one substitutes $u$ with $\bar{\cal D}^2 \psi$ and then varies the action with respect to $\psi$ instead.} $u$ and $\bar{u}$ and substituting the result back into $S_1$ we retrieve the constraint \eqref{v-linear} and the action $S$. If, on the other hand, we vary the auxiliary superfields (now including $v$) we arrive at the dual action
\begin{equation} \label{dual-action}
S' = \int d^4 x {\cal D}^2\bar{\cal D}^2 K(z,\bar{z},u,\bar{u})
\end{equation}
where
\begin{equation}
K(z,\bar{z},u,\bar{u})=F(z,\bar{z},v,\bar{v},t,\bar{t}, \cdots \!, x) - uv- \bar{u}\bar{v} \label{K-pot}
\end{equation}
and
\begin{align}
\frac{\partial F}{\partial v} & = u \label{Legendre-rel1} \\[4pt]
\frac{\partial F}{\partial t}  & = \cdots  = \frac{\partial F}{\partial x} = 0. \label{Legendre-rel2}
\end{align}
If $\eta$ is an ${\cal O}(2)$ multiplet there are no auxiliary fields and $v = \bar{v} = x \in \mathbb{R}$. The equations (\ref{Legendre-rel1}) and (\ref{Legendre-rel2}) are replaced in this case by the single equation
\begin{equation}
\frac{\partial F}{\partial x} = u + \bar{u}. \label{Legendre-rel-O2}
\end{equation}
The equations (\ref{Legendre-rel1})--(\ref{Legendre-rel2}) respectively (\ref{Legendre-rel-O2}) can be implicitly solved to express $v$, $\bar{v}$, $t$, $\bar{t}$, \dots, $x$ respectively $x$ in terms of $z$, $\bar{z}$, $u$, $\bar{u}$.  As an aside, let us note that in the particular case when $G$ is an ${\cal O}(2)$ section over $\A$, the extremization conditions (\ref{Legendre-rel2}) can be reformulated in a unified manner as the vanishing of a contour integral, namely
\begin{equation}
{\cal P}^{B_1\cdots B_{2j-4}} = \oint_{\gamma} \pi_C d\pi^C \pi^{B_1} \cdots \pi^{B_{2j-4}} \frac{\partial G}{\partial \eta} (\eta_{A_1\cdots A_{2j}} \pi^{A_1}\cdots \pi^{A_{2j}}) = 0. \label{extrem-Penrose}
\end{equation}
If $G$ is a function of not one but several multiplets, these equations generalize in a straightforward manner. The Legendre transform relation (\ref{K-pot}) contains additional terms and one has a corresponding set of relations of the type (\ref{Legendre-rel1}) and (\ref{Legendre-rel2}) for each multiplet involved.  In particular, the equation (\ref{extrem-Penrose}) will be replaced by several similar vanishing conditions, one for each ${\cal O}(2j)$ multiplet $\eta$ with $j>1$.

The actions $S$ and $S'$ are dual to each other in the sense that they can be derived from the same first-order action $S_1$. This duality is the superspace equivalent of Hodge duality between $0$ and $2$-form gauge fields in four dimensions. To see this, note that the constraint \eqref{v-linear} can be solved by
\begin{equation}
v = {\cal D}^{\alpha}W_{\alpha} + \bar{\cal D}^{\dot{\alpha}}\bar{W}_{\dot{\alpha}},
\end{equation}
with $W_{\alpha}$ a chiral spinor superfield. This form stays invariant under the infinitesimal transformations
\begin{equation} \label{gauge-transf}
\delta W_{\alpha} = \bar{\cal D}^2{\cal D}_{\alpha} V,
\end{equation}
for any vector superfield $V$. The linear superfield $v$, the chiral spinor superfield $W_{\alpha}$ and the vector superfield $V$ count among their components a $3$-form, a $2$-form and a $1$-form, respectively. It is then natural to think of $v$ as the ``field strength'' of $W_{\alpha}$ and regard \eqref{v-linear} as a Bianchi identity and \eqref{gauge-transf} as a gauge transformation. This can be made more precise with the help of superforms \cite{Gates:1983nr}.

In the dual action \eqref{dual-action}, $K$ is a function of chiral superfields and their anti-chiral conjugates, and thus
is geometrically interpreted as a K\"{a}hler potential. In fact, the action $S'$ inherits the $\N = 2$ supersymmetry
of $S$, so $K$ is a K\"{a}hler potential in one complex structure of a hyperk\"{a}hler manifold.  In sum, then, the
equations \eqref{F-pot}, \eqref{K-pot}, \eqref{Legendre-rel1} and \eqref{Legendre-rel2} can be used to construct hyperk\"{a}hler metrics from meromorphic functions. One can arrive at these same results through purely geometrical means \cite{Hitchin:1986ea,
Ivanov:1995cy}.  This is known in the literature as ``the generalized Legendre transform construction''; its natural mathematical setting
is the theory of twistor spaces of hyperk\"{a}hler manifolds.

Before we continue, let us mention for later use two consequences of the above equations. Taking the derivatives of (\ref{K-pot}) with respect to the holomorphic coordinates and imposing afterwards the generalized Legendre relations (\ref{Legendre-rel1})--(\ref{Legendre-rel2}), one gets that
\begin{equation}
\frac{\partial K}{\partial z} = \frac{\partial F}{\partial z} \qquad\mbox{and}\qquad \frac{\partial K}{\partial u} = -v. \label{Kz-Ku}
\end{equation}
On the other hand, the contour-integral form (\ref{F-pot}) implies that the function $F$ satisfies the following set of second order differential equations
{\allowdisplaybreaks
\begin{align}
F_{z\bar{z}} & = - F_{v\bar{v}} = F_{t\bar{t}\,} = \cdots = (-)^j F_{xx} \nonumber \\[4pt]
F_{z\bar{v}} & = - F_{v\bar{t}\,} = \cdots \nonumber \\[4pt]
F_{zt\,}  & = F_{vv} \hspace{24.5pt} \mbox{etc.} \nonumber \\[4pt]
F_{zv} & = F_{vz} \hspace{25pt} \mbox{etc.} \label{diff-eqs}
\end{align}
}From this point on we switch from the analytic superspace language to a differential geometric language.  Thus, superfields become coordinates, chirality translates into holomorphicity and so on, according to the usual dictionary.

\section{Swann bundles from the generalized Legendre transform} \label{SEC:HKC-from-GLT}

\subsection*{The parameter space action}

The reality constraint \eqref{reality-constr} is preserved only by an $SU(2) \times \mathbb{R}^{\times}$ subgroup of the $GL(2,\mathbb{C})$ group of automorphisms of $\mathbb{CP}^1$. (The latter acts on the inhomogeneous coordinates by linear transformations: $\pi^A \longrightarrow \Lambda^A{}_B\pi^B$, $\det \Lambda \neq 0$.) The reality-preserving automorphisms of $\mathbb{CP}^1$ induce  on the parameter space of ${\cal O}(2j)$ sections an $SO(3) \times \mathbb{R}^{\times}$ action. In practice, it is convenient to combine the three generators $L_1$, $L_2$, $L_3$ of the $SO(3)$ action with the scaling generator $L_0$ corresponding to the $\mathbb{R}^{\times}$ action,
\begin{align}
L_{\pm} = \frac{1}{2}(L_1\pm iL_2) && M_{\pm} = \frac{1}{2}(L_0 \pm iL_3),
\end{align}
and work with this equivalent basis in the complexified space of generators. Explicitly,
\begin{align}
L_+   & =  \bar{v}\frac{\partial}{\partial \bar{z}} + 2\bar{t}\frac{\partial}{\partial \bar{v}} + \cdots - (2j-1) v\frac{\partial}{\partial t} \,- 2j z\frac{\partial}{\partial v} \\[2pt]
M_+  & =  \bar{v}\frac{\partial}{\partial \bar{v}} + 2\bar{t}\frac{\partial}{\partial \bar{t}} \,+ \cdots + (2j-1) v\frac{\partial}{\partial v} + 2j z\frac{\partial}{\partial z} .
\end{align}
The minus index counterparts are obtained simply by complex conjugation: $L_-=\overline{L}_+$ and $M_-=\, \overline{\! M}_+$. One can retrieve with ease the explicit form of the generators of the original basis; for example,
\begin{align}
L_0  & = 2j \!\left( z\frac{\partial }{\partial z}+\bar{z}\frac{\partial }{\partial \bar{z}}
+v\frac{\partial }{\partial v}+\bar{v}\frac{\partial }{\partial \bar{v}}
+\cdots  +x\frac{\partial }{\partial x} \right)  \label{scaling-op} \\
iL_3 & =  2j\! \left( z\frac{\partial}{\partial z} - \bar{z}\frac{\partial}{\partial\bar{z}}\right) + 2(j-1)\! \left( v\frac{\partial}{\partial v} - \bar{v}\frac{\partial}{\partial\bar{v}}\right)  + \cdots  \label{L3-op}
\end{align}
and so on. It is straightforward, albeit slightly tedious, to verify that these generators satisfy indeed the quaternionic algebra,
\begin{align}
[L_i,L_j] = 2 \epsilon_{ijk\,}L_k &&
[L_i,L_0] = 0.
\end{align}

Given the existence of this action on the parameter space, a question arises naturally:  What kind of structure does it induce on the hyperk\"{a}hler spaces constructed from these multiplets by means of the generalized Legendre transform?

The answer to this question involves the criterion that we have formulated earlier in a seemingly different context, in relation to the conformality and  $SU(2)$ invariance of analytic superspace action functionals. In what follows we will prove, generalizing a result of \cite{deWit:2001dj}, that if the function $G(\eta_I)$ satisfies the ${\cal O}(2)$ criterion or its extension, then the induced structure is a Swann bundle structure, that is, the hyperk\"ahler space has an $\mathbb{H}^{\times}$-action comprising one conformal homothetic and three isometric generators \cite{MR1096180}.

\subsection*{Constraints on $F$}

The ${\cal O}(2)$ criterion for the analytic superspace action translates for a theory based exclusively on ${\cal O}(2j)$-type multiplets $\eta_I$ into the following two conditions for $G(\eta_I)$:

 $\hspace{-7pt}1)\ $ it depends on the $\zeta$ variable only through $\eta_I$, not explicitly;

 $\hspace{-7pt}2)\ $ it is made up either of terms homogeneous of degree $2$ at a simultaneous rescaling of the ${\cal O}(2j_I)$ sections $\eta_I$ with weight $2j_I$ or of terms of the form $\eta_I\log \eta_J$, for those $\eta_I$ which are ${\cal O}(2)$ sections.

It is straightforward to check that these two requirements on $G$ imply the following differential constraints for $F$:
\begin{equation}
L_3(F) = 0  \qquad\mbox{and}\qquad  L_0(F) = 2 F \label{diff-O2-0}
\end{equation}
respectively, where $L_0 = \sum_I L_0^{(2j_I)}$ and $L_3 = \sum_I L_3^{(2j_I)}$, with each $L_0^{(2j_I)}$ and $L_3^{(2j_I)}$ of the form \eqref{scaling-op}--\eqref{L3-op} and acting on the parameter space of the corresponding section $\eta_I$. In accordance with our previously stated convention we will drop the index $I$ and work instead with a generic ${\cal O}(2j)$ section $\eta$, in which case $L_0$ and $L_3$ are given precisely by \eqref{scaling-op} and \eqref{L3-op}.

What is the physical significance of these constraints? Recalling that $F$ represents an ${\cal N}=1$  superspace action density, the second condition \eqref{diff-O2-0} simply states that the action density must have conformal symmetry. Choosing a local coordinate trivialization on the Riemann sphere projects at the same time the manifestly ${\cal N}=2$ supersymmetric description down to an ${\cal N}=1$ representation. The first condition \eqref{diff-O2-0} states the invariance of the action density under the residual $U(1)$ subgroup of the ${\cal N}=2$ R-symmetry group $SU(2)_R$. To understand why the particular generator $L_3$ and not some other one comes into play, notice that the equation \eqref{locus} is invariant at $\pi^1 \to e^{i\theta} \pi^1$ and $\pi^2 \to e^{-i\theta} \pi^2$. This leaves $\eta_{A_1\cdots A_{2j}}\pi^{A_1} \cdots \pi^{A_{2j}}$ unchanged provided that we transform as well $z \to e^{-2ji\theta} z$, $v \to e^{-2(j-1)i\theta} v$,  $t \to e^{-2(j-2)i\theta} t$, $\cdots$, $x \to x$. The compensating transformation is clearly generated by $L_3$.

\subsection*{The induced $\mathbb{H}^{\times}$ action}

The first step towards establishing the existence of an induced Swann bundle structure is to determine the action that the generators $L_{\pm}$ and $M_{\pm}$ induce on the holomorphic coordinates $z$ and $u$. For $z$ one gets immediately
\begin{align}
L_+(z) & = 0   & M_+(z) &= 2jz \nonumber \\[2pt]
L_-(z)  & = - \frac{\partial K}{\partial u} & M_-(z) & = 0 .
\end{align}
In deriving the form of $L_-(z)$ we resorted to the second equation (\ref{Kz-Ku}). For the holomorphic coordinate $u$, on the other hand, we obtain
{\allowdisplaybreaks
\begin{align}
L_+(u) & = -M_+(\partial_t F) \hspace{1pt} = - \partial_t[M_+(F)-(2j-2)F] \nonumber \\[5pt]
L_-(u)   & = \phantom{+} M_-(\partial_z F) =  \phantom{+}\! \partial_z[M_-(F)]  \nonumber \\[5pt]
M_+(u) & = \phantom{+}  M_+(\partial_v F) = \phantom{+}  \! \partial_v[M_+(F)-(2j-1)F]  \nonumber \\[5pt]
M_-(u) & = \phantom{+}  M_-(\partial_v F) = \phantom{+}  \! \partial_v[M_-(F)-F] .
\end{align}
}To arrive at these expressions, we first set $u=\partial_vF$ according to \eqref{Legendre-rel1} and then, in the first two cases only, used the equations \eqref{diff-eqs} to transform the second derivatives of $F$. The second set of equalities follows directly by commuting the $M_+$ and $M_-$ operators with the derivatives that act on $F$.

So far we have made no use of the ${\cal O}(2)$ criterion. Notice that in the complexified basis, the two equations \eqref{diff-O2-0} read
\begin{equation}
M_+(F) = M_-(F) = F . \label{diff-O2}
\end{equation}
If we take them into account, the preceding set of relations simplifies significantly and becomes
\begin{align}
L_+(u)  & = 0  && M_+(u) = -2(j-1)u \nonumber \\[2pt]
L_-(u)   & =  \frac{\partial K}{\partial z} && M_-(u) = 0 .
\end{align}
To cast them in this form, we have used further the generalized Legendre relations \eqref{Legendre-rel1} and \eqref{Legendre-rel2} as well as the first equation \eqref{Kz-Ku}.

We conclude that when the criterion holds, the action on the moduli space of ${\cal O}(2j)$ sections generated by $L_{\pm}$ and $M_{\pm}$ induces on the dual hyperk\"{a}hler space an action with generating vector fields given by
\begin{align}
X_+ & = -\frac{\partial K}{\partial\bar{u}} \frac{\partial}{\partial\bar{z}} + \frac{\partial K}{\partial\bar{z}} \frac{\partial}{\partial\bar{u}}  \label{X+op} \\[3pt]
Y_+ & = 2j z\frac{\partial}{\partial z} - 2(j-1)u \frac{\partial}{\partial u} \label{Y+op}
\end{align}
and their conjugates $X_-=\, \overline{\!X}_+$ and $Y_-=\overline{Y}_{\!+}$. These are sections of the complexified tangent bundle; just as before, one can project to real components by means of the relations
\begin{align}
X_{\pm} = \frac{1}{2}(X_1\pm iX_2) && Y_{\pm} = \frac{1}{2}(X_0 \pm iX_3) .
\end{align}

The two conditions \eqref{diff-O2-0} can be also used, in conjunction with the two equations (\ref{Kz-Ku}) and the generalized Legendre transform relations (\ref{K-pot})--(\ref{Legendre-rel2}), to show that
\begin{equation}
X_0(K) = 2 K \qquad \mbox{and} \qquad X_i(K) = 0 \label{H*-on-K}
\end{equation}
for $i=1,2,3$.

Before we proceed to show that $X_0$ is a homothetic conformal Killing vector field, let us pause for a moment to take a closer look at these results.

\subsection*{Alternative representations}

The extension to the case of several multiplets $\eta_I$ of type ${\cal O}(2j_I)$ is straightforward. The vector fields \eqref{X+op} and \eqref{Y+op} generalize, with obvious notations, to
\begin{align}
X_+ & = \sum_I \left[-\frac{\partial K}{\partial\bar{u}_I} \frac{\partial}{\partial\bar{z}^I} + \frac{\partial K}{\partial\bar{z}^I} \frac{\partial}{\partial\bar{u}_I} \right] \label{X+many} \\[0pt]
Y_+ & = \sum_I \left[ 2j_I z^I\frac{\partial}{\partial z^I} - 2(j_I-1)u_I \frac{\partial}{\partial u_I} \right] . \label{Y+many}
\end{align}

Let $J_1$, $J_2$, $J_3$ and $\omega_1$, $\omega_2$, $\omega_3$ be the three standard complex respectively K\"ahler structures of the dual hyperk\"ahler manifold. A well-known result in the theory of hyperk\"ahler manifolds states that the complex-valued $2$-forms $\omega^{\pm} = (\omega_1\pm i\omega_2)/2$ are holomorphic and of type $(2,0)$ respectively anti-holomorphic and of type $(0,2)$ with respect to the complex structure $J_3$. Suppose now that $J_3$ corresponds to the holomorphic set of coordinates $z^I$, $u_I$ that the generalized Legendre transform construction yields.
The twistor-theoretic approach to the generalized Legendre transform of Hitchin, Karlhede, Lindstr\"om and Ro\v{c}ek \cite{Hitchin:1986ea} brings into perspective an aspect which is rather obscure in the analytic superspace approach, namely the fact that $z^I$ and $u_I$ are at the same time holomorphic Darboux coordinates for the holomorphic symplectic form $\omega^+$; that is,
\begin{equation}
\omega^+ = \sum_I dz^I \! \wedge du_I .
\end{equation}
Consider the following local holomorphic symplectomorphism
\begin{align}
Z^I  & = (z^I)^{\frac{1}{2} -(j_I-1)} (u_I)^{\frac{1}{2}-j_I} \nonumber \\[4pt]
U_I & = (z^I)^{\frac{1}{2}+(j_I-1)} (u_I)^{\frac{1}{2}+j_I} .
\end{align}
In this new holomorphic coordinate basis one has
\begin{equation}
\omega^+ = \sum_I dZ^I \! \wedge dU_I \label{Darboux}
\end{equation}
while the vector fields \eqref{X+many} and \eqref{Y+many} take the form
\begin{align}
X_+ & = \sum_I \left[-\frac{\partial K}{\partial\bar{U}_I} \frac{\partial}{\partial\bar{Z}^I} + \frac{\partial K}{\partial\bar{Z}^I} \frac{\partial}{\partial\bar{U}_I} \right] \label{X+alt} \\[0pt]
Y_+ & = \sum_I \left[ Z^I\frac{\partial}{\partial Z^I} + U_I \frac{\partial}{\partial U_I} \right] \label{Y+alt}.
\end{align}
Using only the invariance properties \eqref{H*-on-K} of $K$ and the Darboux property \eqref{Darboux} of $\omega^+$ \footnote{More precisely, for any hyperk\"{a}hler space, in a  coordinate basis holomorphic with respect to a preferred complex structure $J_3$, the components of its holomorphic and anti-holomorphic symplectic forms satisfy the relations
$\omega^+{}_{\mu\rho} \, \omega^{-\rho\nu} = - \delta_\mu{}^\nu$ and
$\omega^{-\rho\sigma} K_{\rho\bar{\mu}}  K_{\sigma\bar{\nu}} = \omega^-{}_{\bar{\mu}\bar{\nu}}$.
Given the particular form \eqref{Darboux} of $\omega^+$ ($\omega^-$ is obtained by complex conjugation),  one can derive a set of relations between the components of the inverse metric and the metric itself. It is these consequences of the Darboux property that one actually employs.} one can show by means of a direct calculation that these generators satisfy the quaternionic algebra
\begin{align}
[X_i,X_j] = 2 \epsilon_{ijk\,}X_k &&
[X_i,X_0] = 0 . \label{H-algebra}
\end{align}
This shows that this form of the generating vector fields is completely general and holds, in the given circumstances, for all Swann bundles, whether constructed from ${\cal O}(2j)$ multiplets or not.

One can use in practice these explicit expressions of the generators to coordinatize the orbits of the $\mathbb{H}^{\times}$ action by setting
\begin{align}
X_+ & = - U \frac{\partial}{\partial\bar{Z}} + Z \frac{\partial}{\partial\bar{U}}  \label{X+UZ} \\[2pt]
Y_+ & = \phantom{+} Z\frac{\partial}{\partial Z} + U \frac{\partial}{\partial U} \label{Y+UZ}
\end{align}
which satisfy automatically the algebra \eqref{H-algebra}. By acting with $X_+$ and $Y_-$ on $Z^I$, $U_I$ and using the two alternative expressions \eqref{X+alt}--\eqref{Y+alt} and \eqref{X+UZ}--\eqref{Y+UZ}, we get that
\begin{align}
\frac{\partial Z^I}{\partial\bar{Z}} & = \frac{\partial Z^I}{\partial\bar{U}} = 0 \\[2pt]
\frac{\partial U_I}{\partial\bar{Z}} & = \frac{\partial U_I}{\partial\bar{U}}\hspace{1pt} = 0
\end{align}
that is, the $Z^I$, $U_I$ coordinates (and consequently the original generalized Legendre transform coordinates $z^I$, $u_I$) depend holomorphically on the fiber coordinates $Z$, $U$.

\subsection*{The homothety}

Clearly, the vector fields $Y_+$ and $\overline{Y}_{\!+}$ are explicitly holomorphic respectively anti-holomorphic with respect to the manifest complex structure associated to the coordinates $z^I$ and $u_I$.  Given that $X_0 = Y_+ + \overline{Y}_{\!+}$ and denoting with $\mu$, $\nu$ the indices holomorphic with respect to this complex structure, we have the following succession of equalities
\begin{equation}
X_0^{\mu}K_{\mu\bar{\nu}} = \partial_{\bar{\nu}} (X_0^{\mu}\partial_{\mu}K) =  \partial_{\bar{\nu}} [Y_+(K)] = \partial_{\bar{\nu}} K.
\end{equation}
In the last step, we made use of the equations \eqref{H*-on-K} to show that $Y_+(K) = K$. Hence, $X_0$ is a gradient vector field:
\begin{equation}
X_0^{\mu} = K^{\mu\bar{\nu}}\partial_{\bar{\nu}} K. \label{X-gradient}
\end{equation}
Denoting with $\nabla$ the Levi-Civita connection associated to the metric $K_{\mu\bar{\nu}}$, this implies that $\nabla_{\nu} X_0^{\mu} = \delta^{\mu}{}_{\nu}$. On the other hand, we also have that $\nabla_{\bar{\nu}} X_0^{\mu} = 0$, an immediate consequence of the holomorphicity structure of $X_0$ and Hermiticity of the connection. Together, these two properties indicate that $X_0$ is a homothetic conformal vector, namely,
\begin{equation}
\nabla X_0 = \mathbb{I}, \label{homothety}
\end{equation}
where $\mathbb{I}$ stands for the identity endomorphism on the tangent bundle. 

\section{Swann bundles: generic considerations} \label{SEC:HKC}

Let $({\cal M},g,\vec{J}\,)$ be a hyperk\"{a}hler variety with a homothetic conformal Killing vector field $X_0$ satisfying the condition (\ref{homothety}). We want to show, based on these assumptions alone, that such a variety is automatically endowed with a Swann bundle structure, that is, with an $\mathbb{H}^{\times}$-action consisting of one conformal homothetic and three isometric generators.

Consider the three vector fields defined by
\begin{equation}
X_i = -J_i X_0 \label{Xi's},
\end{equation}
with $i=1,2,3$. In the context of the generalized Legendre construction discussed above it can be shown that this definition yields precisely the vector fields designated by the same symbols. We shall demonstrate that  $X_0$ together with $X_1$, $X_2$, $X_3$ generate on ${\cal M}$ an  $\mathbb{H}^{\times}$-action with the required properties.

Given a system of coordinates $\{x^{\alpha}\}$ on ${\cal M}$, the the conformal homothetic condition (\ref{homothety}) reads, in components,
\begin{equation}
\nabla_{\beta} X_0^{\alpha} = \delta^{\alpha}{}_{\beta}. \label{nabla-X}
\end{equation}
On the other hand, equation (\ref{Xi's}) together with (\ref{nabla-X}) imply that
\begin{equation}
\nabla_{\beta} X_i^{\alpha} = -J_i^{\,\alpha}{}_{\beta}. \label{nabla-Xi}
\end{equation}
The Lie derivatives of the metric along these vector fields  are evaluated as follows
 \begin{align}
({\cal L}_{X_0}g)_{\alpha\beta} & = X_0^{\gamma}(\nabla_{\gamma}g_{\alpha\beta}) + (\nabla_{\alpha}X_0^{\gamma})g_{\gamma\beta} + (\nabla_{\beta}X_0^{\gamma})g_{\alpha\gamma}
= 2 g_{\alpha\beta} \\[6pt]
({\cal L}_{X_i}g)_{\alpha\beta} & = X_i^{\gamma}(\nabla_{\gamma}g_{\alpha\beta}) + (\nabla_{\alpha}X_i^{\gamma})g_{\gamma\beta} + (\nabla_{\beta}X_i^{\gamma})g_{\alpha\gamma}
= \omega_{i\,\alpha\beta}+\omega_{i\,\beta\alpha}
= 0. \label{L_Xig}
\end{align}
This shows that the action of $X_0$ is conformal while the actions of $X_1$, $X_2$ and $X_3$ are isometric. The partial derivatives may be replaced by covariant derivatives since the Levi-Civita connection is torsion-free. The second set of equalities follows from the equations (\ref{nabla-X}) and (\ref{nabla-Xi}) and the fact that the Levi-Civita connection preserves the metric, i.e. $\nabla g = 0$. In (\ref{L_Xig}), the $\omega_i$ are the K\"{a}hler 2-forms corresponding to the complex structures $J_i$.

We proceed similarly to evaluate their action on the three standard hyperk\"{a}hler complex structures
\begin{align}
({\cal L}_{X_0}J_j)^{\alpha}{}_{\beta} & = X_0^{\gamma}(\nabla_{\gamma} J_j^{\,\alpha}{}_{\beta}) - (\nabla_{\gamma}X_0^{\alpha}) J_j^{\,\gamma}{}_{\beta} + (\nabla_{\beta}X_0^{\gamma}) J_j^{\,\alpha}{}_{\gamma}
= 0 \\[6pt]
({\cal L}_{X_i}J_j)^{\alpha}{}_{\beta} & = X_i^{\gamma}(\nabla_{\gamma} J_j^{\,\alpha}{}_{\beta}) - (\nabla_{\gamma}X_i^{\alpha}) J_j^{\,\gamma}{}_{\beta} + (\nabla_{\beta}X_i^{\gamma}) J_j^{\,\alpha}{}_{\gamma}
= [J_i,J_j]^{\alpha}{}_{\beta}
= 2\epsilon_{ijk\,}J_k^{\,\alpha}{}_{\beta}. \label{L_XiJj}
\end{align}
Thus, the vector field $X_0$ preserves the complex structures whereas the vector fields $X_i$ rotate them into one another.  The second set of equalities follows again from the equations (\ref{nabla-X}) and (\ref{nabla-Xi}) as well as from the compatibility of  the hyperk\"{a}hler complex structures with the the metric connection, i.e. $\nabla J_j = 0$. In (\ref{L_XiJj}) we use moreover the fact that the complex structures satisfy the quaternionic algebra.

Eventually, based on the above results, we have
\begin{align}
[X_0,X_j]  & =  - {\cal L}_{X_0} (J_jX_0) =   -({\cal L}_{X_0}J_j)X_0 - J_j({\cal L}_{X_0}X_0) = 0 \\[5pt]
[X_i,X_j]  & =  - {\cal L}_{X_i}(J_jX_0) \hspace{1pt} = -({\cal L}_{X_i}J_j)X_0 - J_j({\cal L}_{X_i}X_0) \hspace{2pt} = 2\epsilon_{ijk\,}X_k.
\end{align}
This completes the proof of our assertion.

A distinguishing feature of Swann bundles among hyperk\"{a}hler manifolds is the fact that there exists a scalar function $\chi:{\cal M} \rightarrow \mathbb{R}$, defined up to the addition of a constant, which is simultaneously a K\"{a}hler potential for $J_1$, $J_2$ and $J_3$. To see that, let us first observe that the homothetic Killing vector condition (\ref{nabla-X}) implies, based on the compatibility of the Levi-Civita connection with the metric, that $\nabla_{\beta}(X_0)_{\alpha} = g_{\alpha\beta}$, where $(X_0)_{\alpha} = g_{\alpha\beta}X_0^{\beta}$ are the components of the 1-form dual to $X_0$ with respect to the metric bilinear form. Exploiting further the symmetry of the metric  and the torsion-free character of the connection one derives the closure condition $\partial_{\alpha}(X_0)_{\beta} - \partial_{\beta} (X_0)_{\alpha} = 0$.
This implies that, at least locally, the dual 1-form can be expressed as  the total derivative of a scalar function $\chi$, and so $X_0$ must be a gradient vector field,
\begin{equation}
X_0^{\alpha} = g^{\alpha\beta}\partial_{\beta}\chi \label{gradient}
\end{equation}
where, as usual, $g^{\alpha\beta}$ represents the inverse metric.

Choose now an arbitrary complex structure $J_k$ and define the complex-valued 1-form
\begin{equation}
\theta_k = (X_0^{\alpha} + i X_k^{\alpha})g_{\alpha\beta}\,dx^{\beta}. \label{theta_k}
\end{equation}
On one hand, based on (\ref{gradient}), we can write this as follows
\begin{equation}
\theta_k = \partial_{\alpha}\chi(\delta^{\alpha}{}_{\beta} + i J_k^{\,\alpha}{}_{\beta}) dx^{\beta} = 2\, \bar{\partial}_{J_k} \chi. \label{tk}
\end{equation}
On the other hand, we have
\begin{equation}
d\theta_k = \nabla_{\alpha}\theta_{k\beta}\, dx^{\alpha} \wedge dx^{\beta} = 2i\, \omega_k \label{d-theta_k}
\end{equation}
where $\omega_k$ is the K\"{a}hler 2-form corresponding to the complex structure $J_k$. The second equality follows by substituting (\ref{theta_k}) and then making use of the properties (\ref{nabla-X}) and (\ref{nabla-Xi}). The exterior derivative decomposes along holomorphic lines into $d=\partial_{J_k} + \bar{\partial}_{J_k}$, hence from (\ref{tk}) and (\ref{d-theta_k}) we infer that
\begin{equation}
\omega_k = -i\, \partial_{J_k}\bar{\partial}_{J_k} \chi
\end{equation}
which means that $\chi$ is a K\"{a}hler potential for $\omega_k$. Since our choice of complex structure was arbitrary, it follows that the function $\chi$ is equally a K\"{a}hler potential for $J_1$, $J_2$ and $J_3$.

The function $\chi$ is $SU(2)$-invariant and scales with weight 2 under the action of $X_0$. To see this, note first that the equations (\ref{nabla-X}) and (\ref{gradient}) can be used to show that
\begin{equation}
g_{\alpha\beta} = \nabla_{\alpha}\nabla_{\beta}\chi.
\end{equation}
Following \cite{Gibbons:1998xa}, let $V = g^{\alpha\beta}\partial_{\alpha}\chi\partial_{\beta}\chi$. Based on the previous relation, one has $\partial_{\alpha}V = 2\,\partial_{\alpha}\chi$. We can choose the arbitrary integration constant such that $V = 2\chi$. In this case, the first equation can be re-written in the form
\begin{equation}
X_0(\chi) = 2\chi.
\end{equation}
So $\chi$ is an eigenvalue of the homothety generator, $X_0$. From this and (\ref{gradient}) one infers then immediately that
\begin{equation}
\hspace{-11pt} X_i(\chi) = 0
\end{equation}
for $i=1,2,3$. Comparing these equations with \eqref{H*-on-K}, we conclude that, up to a constant, $\chi=K$ and the generalized Legendre transform construction yields, in the case of Swann bundles, not just any K\"{a}hler potential corresponding to the manifest complex structure but in fact the hyperk\"{a}hler potential itself.

\section{Applications} \label{SEC:App}

We conclude with a brief and selective review of several generalized Legendre transform constructions of Swann bundles discussed in the literature.

The $F$-potential
\begin{equation}
F = \frac{1}{2\pi i} \oint_{\gamma} \frac{d\zeta}{\zeta} \eta^{(2)} \log \eta^{(2)} = \int^{\ \alpha}_{-1/\bar{\alpha}} \frac{d\zeta}{\zeta} \eta^{(2)}
\end{equation}
with $\gamma$ a closed eight-shaped contour surrounding the roots $\alpha$ and $-1/\bar{\alpha}$ of $\eta^{(2)}$ was shown in \cite{Karlhede:1984vr} to generate the standard flat metric on $\mathbb{H}^{\times} \simeq \mathbb{R}^4\setminus\! \{0\}$. This can be viewed as a trivial Swann bundle over a point. The monodromy of the logarithm can be used to turn the closed-contour complex integral into an open-contour one. The $G$-function corresponding to the closed-contour formulation is not an ${\cal O}(2)$ section over $\mathbb{CP}^1$ but rather an affine ${\cal O}(2)$ section; nevertheless, both conditions \eqref{diff-O2-0} are fully satisfied in this case. In the alternative, open-contour formulation, the corresponding $G$-function is clearly an ${\cal O}(2)$ section and moreover, the endpoints of the contour transform according to the standard action of $SU(2)$ on $\mathbb{CP}^1$. The requirements of our criterion are thus satisfied, either way.

The related $F$-potential
\begin{equation}
F = -\frac{1}{2\pi i} \oint_{\gamma_0} \frac{d\zeta}{\zeta} (\eta^{(2)})^2  + \frac{1}{2\pi i} \oint_{\gamma} \frac{d\zeta}{\zeta} \eta^{(2)} \log \eta^{(2)} ,
\end{equation}
with $\gamma_0$ denoting a closed loop around $\zeta = 0$ and $\gamma$ the same contour as above, generates the Taub-NUT metric \cite{Karlhede:1984vr}. The corresponding $G$-function does not satisfy the requirements of the ${\cal O}(2)$ criterion, due to its first term, which is an ${\cal O}(4)$ section.  The first equation \eqref{diff-O2-0} still holds in this case, but the second one does not. One can still ask the question what structure, if any, does the natural $SO(3) \times \mathbb{R}^{\times}$ action on the parameter space of the $\eta^{(2)}$ sections induce on the resulting hyperk\"ahler space?   An analysis along the lines of the first part of section \ref{SEC:HKC-from-GLT} (for some details see e.g. \cite{MM1}) shows that in this case the scaling isometry is broken, but the $SO(3)$ piece of this action survives and induces a non-triholomorphic $SO(3)$ isometry that rotates the sphere of complex structures, a well-known feature of the Taub-NUT space. Of course, besides this isometry, one has a further triholomorphic $U(1)$ isometry, as for any ${\cal O}(2)$-based generalized Legendre transform construction.

$F$-potentials constructed out of a number of $n+1$ ${\cal O}(2)$ multiplets, of the type
\begin{equation}
F = \frac{1}{2\pi i} \oint \frac{d\zeta}{\zeta} \frac{{\cal F}(\eta^{(2)}_1, \cdots,\eta^{(2)}_n)}{\eta^{(2)}_0} ,
\end{equation}
with ${\cal F}$ a meromorphic function of $n$ variables homogeneous of degree $2$, have been used in \cite{Rocek:2005ij,Rocek:2006xb,Neitzke:2007ke} to describe the Swann bundles endowed with $n+1$ commuting $U(1)$ isometries associated to quaternionic K\"ahler spaces obtained via the c-map of \cite{Cecotti:1988qn,Ferrara:1989ik} from the projective (non-rigid) special K\"ahler manifolds with meromorphic prepotential ${\cal F}$. The c-map construction corresponds to the dimensional reduction of ${\cal N} = 2$ supergravity coupled to vector multiplets from dimension four to dimension three, followed by the dualization of the vector multiplets into hypermultiplets.

Higher-order multiplets have been employed as well to construct Swann bundle metrics. The ${\cal O}(2)\oplus {\cal O}(4)$ $F$-potential
\begin{equation}
F =  \oint_{\gamma} \frac{d\zeta}{\zeta} \left[ \frac{(\eta^{(2)})^2}{\sqrt{\eta^{(4)}}} - \sqrt{\eta^{(4)}} \right],
\end{equation}
with the contour $\gamma$ surrounding the branch-cuts of $\sqrt{\eta^{(4)}}$ in such a way as to guarantee a real outcome for the integrals, was conjectured in \cite{Anguelova:2004sj}, based on symmetry  arguments and asymptotic behavior, to describe the nonperturbative universal hypermultiplet moduli space metric due to five-brane instantons.

\vskip30pt
\noindent {\large \bf Acknowledgements} \\ [10pt]
The authors wish to thank Martin Ro\v{c}ek for support and valuable discussions.
AN was supported in part by the Martin A.\ and Helen
Chooljian Membership at the Institute for Advanced Study, and in part by the
NSF under grant numbers PHY-0503584 and PHY-0804450.

\bibliographystyle{utphys}
\bibliography{tensor-conformal}

\end{document}